\documentclass[preprint,12pt]{elsarticle}

\usepackage{lineno,hyperref}
\usepackage{amsmath}
\usepackage{amssymb}
\newcommand{\haf}{\frac{1}{2}}
\newcommand{\la}{{\langle}}
\newcommand{\ra}{{\rangle}}
\newcommand{\p}{{\partial}}
\modulolinenumbers[5]

\journal{Physics Letters A}

\begin{document}

\begin{frontmatter}

\title{Exact Density matrix of an oscillator-bath system: Alternative derivation}

\author{Fardin Kheirandish}
\address{Department of Physics, Faculty of Science, University of Kurdistan, P.O.Box 66177-15175, Sanandaj, Iran}

\begin{abstract}
\noindent Starting from a total Lagrangian describing an oscillator-bath system, an alternative derivation of exact quantum propagator is presented. Having the quantum propagator, the exact density matrix, reduced density matrix of the main oscillator and thermal equilibrium fixed point are obtained. The modified quantum propagator is obtained in the generalised  case where the main oscillator is under the influence of a classical external force. By introducing auxiliary classical external fields, the generalised quantum propagator or generating functional of position correlation functions is obtained.
\end{abstract}

\begin{keyword}
\texttt{Density matrix}\sep Oscillator-Bath\sep Propagator \sep Generating function
\end{keyword}

\end{frontmatter}

\linenumbers

\section{Introduction}
%%%%%%%%%%%%%%%%%%%%%%%%%%%%%%%%%%%%%%%%%%%%%%%%%%%%%%%%%%%%%%%%%%%%%%%%%%%%%%%%%%%%%%%%%%%%%%%%%%%%%%%%%%%%%%%%%%%%%%%%%%%%%%%%%%%%%%%%%%
\noindent The quantum propagator is the most important function in quantum theories \cite{Propagator-1,Propagator-2}. Knowing the quantum propagator, we can obtain all measurable quantities related to the physical system exactly, that is we have a complete physical description of the underline system in any time. Unfortunately, except for some simple physical systems, obtaining the exact form of quantum propagator is usually a difficult task and we have to invoke perturbative methods. Among different approaches to find quantum propagatorو we can refer to two main approaches. In the first method, quantum propagator is written as a bilinear function in eigenvectors of the Schr\"{o}dinger equation. The main task in this method is to find the eigenfunctions of the Hamiltonian which are usually difficult to find and even having these eigenfunctions, extracting a closed form quantum propagator from them may be cumbersome. The second approach is based on the Feynman path integral technique \cite{Feynman-1,Kleinert,Zin}. One of the most efficient features of this method is its perturbative technique known as Feynman diagrams which extends the applicability of the method to the era of non-quadratic Lagrangians. The path integral technique has been applied to oscillator-bath system in \cite{Path-0,Path-1,Path-2,Path-3,Path-4,Path-5,Path-6}.

Here we follow an alternative approach to find the quantum propagator. This approach which we will describe in detail is based on the position and momentum operators in Heisenberg picture. In this scheme, using elementary quantum mechanical relations, two independent partial differential equations are found that quantum propagator satisfy in. The solutions of these partial differential equations are easily found and unknown functions are determined from basic properties of quantum propagators. The first message of the present paper is that this method compared to other methods to derive the quantum propagator of an oscillator-bath system with a linear coupling is easier to apply and in particular, comparing with path integral technique, there is no need to introduce more advanced mathematical notions like infinite integrations, operator determinant and Weyl ordering. The second message is that since we will find a closed form for the total quantum propagator, we will find a closed form density matrix describing the combined oscillator-bath system. Also, by tracing out the bath degrees of freedom, we find a reduced density matrix describing the main oscillator in any time. In the following, we will generalise the oscillator-bath model by including external classical sources in Hamiltonian, and find the modified quantum propagator under the influence of classical forces. The modified quantum propagator can be interpreted also as a generating functional from which time-ordered correlation functions among different position operators can be determined \cite{Greiner}. The basic ingredient of the approach is a symmetric time-independent matrix $B$, (Eq.(\ref{14-2}) depending on natural frequencies of the bath oscillators and coupling constants. Therefore, from numerical or simulation point of view, the only challenge is finding the inverse of the matrix $B$ or equivalently diagonalizing it.

The efficiency of the method introduced here in determining the exact form of the quantum propagator for quadratic Lagrangians, inspires the idea of developing a perturbative approach to include non-quadratic Lagrangians too. The process presented to determine the quantum propagator, suggest that these perturbative techniques may be based on perturbative solutions of nonlinear partial differential equations. This development deserves to be investigated in an independent work.
%%%%%%%%%%%%%%%%%%%%%%%%%%%%%%%%%%%%%%%%%%%%%%%%%%%%%%%%%%%%%%%%%%%%%%%%%%%%%%%%%%%%%%%%%%%%%%%%%%%%%%%%%%%%%%%%%%%%%%%%%%%%%%%%%%%%%%%%%%
\section{Lagrangian}
%%%%%%%%%%%%%%%%%%%%%%%%%%%%%%%%%%%%%%%%%%%%%%%%%%%%%%%%%%%%%%%%%%%%%%%%%%%%%%%%%%%%%%%%%%%%%%%%%%%%%%%%%%%%%%%%%%%%%%%%%%%%%%%%%%%%%%%%%%
\noindent In this section, we set the stage for what will be investigated in the following sections. We start with a total Lagrangian describing an interacting oscillator-bath system. Then from the corresponding Hamiltonian and Heisenberg equations of motion, we find explicit expressions for position and momentum operators as the main ingredients of an approach that will be applied in the next section. The Lagrangian describing a main oscillator interacting linearly with a bath of oscillators is given by \cite{Weiss}
\begin{equation}\label{1}
  L=\haf \dot{x}^2-\haf\omega_0^2 x^2+\sum_{i=1}^N \haf (\dot{X}^2_i-\omega_i^2 X^2_i)+\sum_{i=1}^N g_i X_i x,
\end{equation}
Eq.(\ref{1}) can be rewritten in a more compact form as
\begin{equation}\label{2}
  L=\haf \sum_{\mu=0}^N (\dot{Y}^2_\mu-\omega_\mu^2 Y_\mu^2)+\haf\sum_{\mu,\nu=0}^N Y_\mu \Omega_{\mu\nu}^2 Y_\nu,
\end{equation}
where the matrix $\Omega^2_{\mu\nu}$ is given by
\begin{equation}\label{3}
  \Omega^2_{\mu\nu}=\left(
                    \begin{array}{ccccc}
                      0 & g_1 & g_2 & \cdots & g_N \\
                      g_1 & 0 & 0 & \cdots & 0 \\
                      g_2 & 0 & 0 & \cdots & 0 \\
                      \vdots & \vdots & \vdots & \ddots & \vdots \\
                      g_N & 0 & 0 & \cdots & 0 \\
                    \end{array}
                  \right),
\end{equation}
and
\begin{equation}\label{4}
  Y_0=x,\,\,\,\,Y_k=X_k,\,\,\,\,k=1,\cdots,N.
\end{equation}
The corresponding Hamiltonian is
\begin{equation}\label{5}
  H=\haf \sum_{\mu=0}^N (P^2_\mu+\omega_\mu^2 Y_\mu^2)-\haf\sum_{\mu,\nu=0}^N Y_\mu \Omega_{\mu\nu}^2 Y_\nu,
\end{equation}
where $P_\mu=\dot{Y}_\mu$ is the canonical conjugate momentum corresponding to the canonical position $Y_\mu$. The system is quantized by imposing the equal-time commutation relations
\begin{eqnarray}\label{6}
&&  [\hat{Y}_\mu, \hat{P}_\nu] = i\hbar\,\delta_{\mu\nu},\nonumber\\
&&  [\hat{Y}_\mu, \hat{Y}_\nu] = [\hat{P}_\mu, \hat{P}_\nu]=0,
\end{eqnarray}
and from Heisenberg equations of motion one finds
\begin{equation}\label{7}
 \ddot{\hat{Y}}_\mu +\omega_\mu^2 \hat{Y}_\mu=\sum_\nu \Omega_{\mu\nu}^2 \hat{Y}_\nu.
\end{equation}
Note that $(\hat{Y}_0, \hat{P}_0)$ refer to the position and momentum of the main oscillator and $(\hat{Y}_k, \hat{P}_k),\,\,(k=1,\cdots,N)$ refer to position and momentum operators of bath oscillators.
Taking the Laplace transform from both sides of Eq.(\ref{7}) we find
\begin{equation}\label{8}
  \sum_{\nu}\Lambda_{\mu\nu}(s)\hat{\tilde{Y}}_\mu (s)=s \hat{Y}_\mu (0)+\hat{P}_\mu (0),
\end{equation}
where the $N+1$-dimensional matrix $\Lambda$ is defined by
\begin{equation}\label{9}
  \Lambda_{\mu\nu} (s)=[(s^2 +\omega_\mu^2)\delta_{\mu\nu}-\Omega_{\mu\nu}^2].
\end{equation}
Therefore, applying the inverse matrix, we find
\begin{equation}\label{10}
  \hat{\tilde{Y}}_\mu (s)=\sum_{\nu} [s\Lambda^{-1}_{\mu\nu} (s)\hat{Y}_\nu (0)+\Lambda^{-1}_{\mu\nu} (s)\hat{P}_\nu (0)],
\end{equation}
and a formal solution is obtained by inverse Laplace transform as
\begin{equation}\label{11}
  \hat{Y}_\mu (t)=\dot{F}_{\mu\nu} (t) \hat{Y}_\nu (0)+F_{\mu\nu} (t) \hat{P}_\nu (0),
\end{equation}
where we defined
\begin{equation}\label{12}
  F_{\mu\nu} (t)=\mathcal{L}^{-1}[\Lambda^{-1}(s)]_{\mu\nu}.
\end{equation}
The matrix $\Lambda$ is explicitly given by
\begin{equation}\label{13}
  \Lambda (s)=\left(
                    \begin{array}{ccccc}
                      s^2 + \omega_0^2 & -g_1 & -g_2 & \cdots & -g_N \\
                      -g_1 & s^2 + \omega_1^2 & 0 & \cdots & 0 \\
                      -g_2 & 0 & s^2 + \omega_2^2 & \cdots & 0 \\
                      \vdots & \vdots & \vdots & \ddots & \vdots \\
                      -g_N & 0 & 0 & \cdots & s^2 + \omega_N^2 \\
                    \end{array}
                  \right),
\end{equation}
which can be rewritten as
\begin{equation}\label{14-1}
  \Lambda (s)=s^2 \,\mathbb{I}+B,
\end{equation}
wherein
\begin{equation}\label{14-2}
 B=\left(
                    \begin{array}{ccccc}
                      \omega_0^2 & -g_1 & -g_2 & \cdots & -g_N \\
                      -g_1 & \omega_1^2 & 0 & \cdots & 0 \\
                      -g_2 & 0 & \omega_2^2 & \cdots & 0 \\
                      \vdots & \vdots & \vdots & \ddots & \vdots \\
                      -g_N & 0 & 0 & \cdots & \omega_N^2 \\
                    \end{array}
                  \right).
\end{equation}
The inverse matrix can be formally written as
\begin{eqnarray}\label{14-3}
  \Lambda^{-1} (s) &=& \frac{1}{s^2 \,\mathbb{I}+B}=\frac{1}{s^2}\,\frac{1}{\mathbb{I}+\frac{1}{s^2}\,B}\nonumber \\
   &=& \frac{1}{s^2}\,\bigg(\mathbb{I}-\frac{1}{s^2}\,B+\frac{1}{s^4}B^2-\cdots\bigg)\nonumber \\
   &=& \sum_{n=0}^\infty \frac{(-1)^n}{s^{2n+2}}\,B^n,\,\,\,\,(B^0=\mathbb{I}).
\end{eqnarray}
Therefore, from Eq.(\ref{12}) we have
\begin{eqnarray}\label{15}
 && F_{\mu\nu}(t)=\sum_{n=0}^\infty \frac{(-1)^n \,t^{2n+1}}{(2n+1)!}\,(B^n)_{\mu\nu},\nonumber\\
 && \dot{F}_{\mu\nu}(t)=\Big(\frac{dF}{dt}\Big)_{\mu\nu}=\sum_{n=0}^\infty \frac{(-1)^n \,t^{2n}}{(2n)!}\,(B^n)_{\mu\nu}.
\end{eqnarray}
The equations Eqs.(\ref{15}) can be formally written as
\begin{eqnarray}\label{15-1}
 && F(t)=\frac{1}{\sqrt{B}}\,\sin(\sqrt{B}\,t),\nonumber\\
 && \dot{F}(t)=\cos(\sqrt{B}\,t).
\end{eqnarray}
From Eqs.(\ref{15}) we deduce that the matrices $F_{\mu\nu}(t)$ and $\dot{F}_{\mu\nu}(t)$ are odd and even in $t$, respectively.
\subsection{Connection to the previous works}
\noindent The Eq. (\ref{11}) has been appeared in \cite{Haake} with a minor change of notation in the framework of Ullersma diagonalisation technique \cite{Ullersma}. Let the
matrix $X_{\mu\nu}$ be a unitary matrix that diagonalizes the orthogonal matrix $B$ given by Eq. (\ref{14-2}) with corresponding eigenvalues $z^2_{\alpha},\,(\alpha=0,1,\cdots,N)$. Therefore, in matrix notation we have
\begin{equation}\label{C1}
 (X^t B X)_{\alpha\beta}=z^2_{\alpha}\,\delta_{\alpha\beta},
\end{equation}
and using the first equation of Eq. (18), we find \cite{Haake}
\begin{equation}\label{C2}
F_{\mu\nu} (t)=\sum_{\alpha=0}^N X_{\mu\alpha}X_{\nu\alpha}\frac{1}{z_{\alpha}}\sin z_{\alpha}t.
\end{equation}
The eigenvalues $z^2_{\alpha}$ of the matrix $B$, satisfy the characteristic equation
\begin{equation}\label{C3}
\det(B-z^2\mathbb{I})=0\Rightarrow \left|
                    \begin{array}{ccccc}
                      \omega_0^2-z^2 & -g_1 & -g_2 & \cdots & -g_N \\
                      -g_1 & \omega_1^2-z^2 & 0 & \cdots & 0 \\
                      -g_2 & 0 & \omega_2^2 & \cdots & 0 \\
                      \vdots & \vdots & \vdots & \ddots & \vdots \\
                      -g_N & 0 & 0 & \cdots & \omega_N^2-z^2 \\
                    \end{array}
                  \right|=0,
\end{equation}
the determinant can be evaluated using the mathematical induction leading to the following characteristic equation
\begin{equation}\label{C4}
  g(z)=z^2-\omega_0^2-\sum_{n}\frac{g_n^2}{z^2-\omega_n^2}=0.
\end{equation}
By making use of Eq. (\ref{7}), we find the following quantum Langevin equation for the main oscillator
\begin{equation}\label{C5}
\ddot{\hat{Y}}_0 (t)-\int_0^t dt'\,\chi(t-t')\,\hat{Y}_0 (t')+\omega_0^2 \,Y_0 (t)=\Upsilon(t),
\end{equation}
where the susceptibility of the environment is defined by
\begin{equation}\label{C6}
\chi(t)=\sum_{k=1}^N g^2_k\,\frac{\sin(\omega_k t)}{\omega_k},
\end{equation}
and the noise operator by
\begin{equation}\label{C7}
\hat{\Upsilon}(t)=\sum_{k=1}^N g_k \,\big[\cos(\omega_k t)\hat{Y}_k (0)+\frac{\sin(\omega_k t)}{\omega_k}\hat{P}_k (0)\big],
\end{equation}
where $\hat{Y}_k (0)$ and $\hat{P}_k (0)$ are the position and momentum operators at initial time ($t=0$). Taking the Laplace transform of the Langevin equation Eq. (\ref{C5}), we will find the Laplace transform of the corresponding Green's function as
\begin{equation}\label{C8}
\tilde{G}(s)=\frac{1}{s^2-\tilde{\chi}(s)+\omega_0^2}=\frac{1}{s^2+\omega_0^2-\sum\limits_{k=1}^N \frac{g_k^2}{s^2+\omega_k^2}},
\end{equation}
where
\begin{equation}\label{C9}
\tilde{\chi}(s)=\sum_{k=1}^N \frac{g_k^2}{s^2+\omega_k^2}.
\end{equation}
The Green's function in frequency space ($G(\omega)$) can be obtained from the Laplace transformed Green's function $\tilde{G}(s)$ using the identity $G(\omega)=\tilde{G}(i\omega)$, therefore,
\begin{equation}\label{C10}
G(\omega)=\frac{-1}{\omega^2-\omega_0^2-\sum\limits_{k=1}^N \frac{g_k^2}{\omega^2-\omega_k^2}}=\frac{-1}{g(\omega)},
\end{equation}
that is the roots of the characteristic equation $g(z)=0$ are the poles of the Green's function $G(\omega)$ in frequency domain.
%%%%%%%%%%%%%%%%%%%%%%%%%%%%%%%%%%%%%%%%%%%%%%%%%%%%%%%%%%%%%%%%%%%%%%%%%%%%%%%%%%%%%%%%%%%%%%%%%%%%%%%%%%%%%%%%%%%%%%%%%%%%%%%%%%%%%%%%%%
\section{Quantum Propagator}
%%%%%%%%%%%%%%%%%%%%%%%%%%%%%%%%%%%%%%%%%%%%%%%%%%%%%%%%%%%%%%%%%%%%%%%%%%%%%%%%%%%%%%%%%%%%%%%%%%%%%%%%%%%%%%%%%%%%%%%%%%%%%%%%%%%%%%%%%%
\noindent In this section a novel scheme to derive the quantum propagator of the combined oscillator-bath system is introduced in detail. Let $|y_0\ra$ be an eigenket of $\hat{Y}_0$ and $|y_k\ra$ an eigenket of $\hat{Y}_k$, then in Heisenberg picture, we can write
\begin{equation}\label{16}
 \hat{Y}_\mu (t)\,|\mathbf{y},t\ra=y_\mu \,|\mathbf{y},t\ra,
\end{equation}
where for notational simplicity the tensor product is abbreviated as
\begin{equation}\label{17}
  |\mathbf{y},t\ra=|y_0,t\ra\otimes|y_1,t\ra\otimes\cdots\otimes|y_N,t\ra=|y_0,\cdots,y_N,t\ra.
\end{equation}
Multiplying Eq.(\ref{16}) from the left by $\la \mathbf{y}'|$ and using Eq.(\ref{11}), we find
\begin{equation}\label{18}
  \sum_{\nu=0}^N\bigg(\dot{F}_{\mu\nu} (t)\,y'_\nu-i\hbar\,F_{\mu\nu} (t)\,\frac{\p}{\p y'_\nu}\bigg)
  \,\mathcal{K}(\mathbf{y}'|\mathbf{y},t)=y_\mu\,\mathcal{K}(\mathbf{y}'|\mathbf{y},t),
\end{equation}
where we have defined the function $\mathcal{K}$ as
\begin{equation}\label{18-1}
 \la \mathbf{y}'|\mathbf{y},t\ra=\mathcal{K}(\mathbf{y}'|\mathbf{y},t),
\end{equation}
and made use of the identities
\begin{eqnarray}\label{19}
  \la \mathbf{y}'|\hat{Y}_\mu (0) &=& y'_\mu \la \mathbf{y}'|,\nonumber \\
  \la \mathbf{y}'|\hat{P}_\mu (0) &=& -i\hbar\,\frac{\p}{\p y'_\mu}\la \mathbf{y}'|.
\end{eqnarray}
Eq.(\ref{18}) can be rewritten as
\begin{equation}\label{20}
  \sum_{\nu=0}^N F_{\mu\nu} (t)\,\frac{\p}{\p y'_\nu}\ln \mathcal{K}(\mathbf{y}'|\mathbf{y},t)=
  \frac{i}{\hbar}\bigg(y_\mu-\sum_{\nu}\dot{F}_{\mu\nu} (t)\,y'_\nu\bigg).
\end{equation}
The right hand side of Eq.(\ref{20}) is linear in $y'_\mu$, so the following quadratic form can be assumed for $\ln \mathcal{K}$
\begin{equation}\label{21}
  \ln\mathcal{K}(\mathbf{y}'|\mathbf{y},t)=A(\mathbf{y},t)+\sum_{\mu=0}^N A_\mu (\mathbf{y},t)y'_\mu+\haf \sum_{\mu,\nu=0}^N y'_\mu C_{\mu\nu} (\mathbf{y},t) y'_\nu,
\end{equation}
where $C_{\mu\nu}=C_{\nu\mu}$. By inserting Eq.(\ref{21}) into Eq.(\ref{20}), we easily find
\begin{eqnarray}\label{22}
  A_\mu (y,t) &=& \frac{i}{\hbar}\,\sum_{\nu=0}^N F^{-1}_{\mu\nu} (t)\,y_\nu,\nonumber\\
  C_{\mu\nu} (t) &=& -\frac{i}{\hbar}\,\sum_{\sigma=0}^N F^{-1}_{\mu\sigma} (t)\,\dot{F}_{\sigma\nu} (t),
\end{eqnarray}
therefore, in dyadic notation, we can write
\begin{equation}\label{23}
  \mathcal{K}(\mathbf{y}'|\mathbf{y},t)=e^{A(\mathbf{y},t)}e^{\frac{i}{\hbar}\mathbf{y}'\cdot\mathbf{F}^{-1}(t)\cdot \mathbf{y}}
  e^{-\frac{i}{2\hbar}\,\mathbf{y}'\cdot \mathbf{F}^{-1} (t)\dot{\mathbf{F}}(t)\cdot \mathbf{y}'}.
\end{equation}
The form of $A(\mathbf{y},t)$ can be determined from the properties of propagators. Since the Hamiltonian Eq.(\ref{5}) is time-independent, we can write
\begin{equation}\label{24}
  \mathcal{K}(\mathbf{y}'|\mathbf{y},t)=\la \mathbf{y}'|\mathbf{y},t\ra=\la \mathbf{y}'|e^{\frac{it}{\hbar}\hat{H}}|\mathbf{y}\ra.
\end{equation}
Eq.(\ref{24}), is invariant under successive transformations (i) complex conjugation (ii) $\mathbf{y}\leftrightarrow \mathbf{y}'$ (iii) $t\rightarrow -t$, therefore,
\begin{equation}\label{25}
  \mathcal{K}(\mathbf{y}'|\mathbf{y},t)=\mathcal{K}^{*}(\mathbf{y}|\mathbf{y}',-t),
\end{equation}
leading to
\begin{eqnarray}\label{26}
 e^{A(\mathbf{y},t)} &=& e^{\varphi(t)}\,e^{-\frac{i}{2\hbar}\,\mathbf{y}\cdot \mathbf{F}^{-1} (t)\dot{\mathbf{F}}(t)\cdot \mathbf{y}},\nonumber\\
 \varphi^{*} (-t) &=& \varphi(t).
\end{eqnarray}
Note that in Sec.VI, the Hamiltonian will be time-dependent and to find $A(\mathbf{y},t)$ we can not use these transformations and we will follow another approach. Up to now the form of the propagator is as follows
\begin{eqnarray}\label{27}
  \mathcal{K}(\mathbf{y}'|\mathbf{y},t) &=& e^{\varphi(t)}\,e^{-\frac{i}{2\hbar}\,\mathbf{y}\cdot \mathbf{F}^{-1} (t)\dot{\mathbf{F}}(t)\cdot \mathbf{y}}e^{\frac{i}{\hbar}\mathbf{y}'\cdot\mathbf{F}^{-1}(t)\cdot \mathbf{y}}
  e^{-\frac{i}{2\hbar}\,\mathbf{y}'\cdot \mathbf{F}^{-1} (t)\dot{\mathbf{F}}(t)\cdot \mathbf{y}'},\nonumber\\
  &=& e^{\varphi(t)}\,e^{-\frac{i}{2\hbar}[\mathbf{y}\cdot \mathbf{F}^{-1} \dot{\mathbf{F}}\cdot \mathbf{y}+
  \mathbf{y}'\cdot \mathbf{F}^{-1} \dot{\mathbf{F}}\cdot \mathbf{y}'-2\mathbf{y}'\cdot \mathbf{F}^{-1}\cdot \mathbf{y}]}.
\end{eqnarray}
From Eqs.(\ref{15-1}) we find the following asymptotic behaviours of Matrices $\mathbf{F},\,\mathbf{F}^{-1}$, and $\dot{\mathbf{F}}$
\begin{eqnarray}\label{28}
  \lim_{t\rightarrow 0} \mathbf{F}(t) \thickapprox t\,\mathbb{I},\nonumber \\
  \lim_{t\rightarrow 0} \mathbf{F}^{-1}(t) \thickapprox \frac{1}{t}\,\mathbb{I},\nonumber \\
  \lim_{t\rightarrow 0} \dot{\mathbf{F}}(t) \thickapprox \mathbb{I},
\end{eqnarray}
By inserting these asymptotic behaviours into Eq.(\ref{27}) we find
\begin{equation}\label{29}
  \lim_{t\rightarrow 0}\mathcal{K}(\mathbf{y}'|\mathbf{y},t)=\delta(\mathbf{y}'-\mathbf{y})=
  \lim_{t\rightarrow 0} e^{\varphi(t)}\,e^{-\frac{i}{2\hbar t}(\mathbf{y}'-\mathbf{y})^2},
\end{equation}
comparing Eq.(\ref{29}) with the following one-dimensional representation of Dirac delta function
\begin{equation}\label{30}
  \lim_{t\rightarrow 0} \sqrt{\frac{A}{\pi t}}\,e^{-\frac{A}{t}\,(x-x')^2}=\delta(x-x'),
\end{equation}
we deduce immediately
\begin{equation}\label{31}
  \lim_{t\rightarrow 0} e^{\varphi(t)}=\bigg(\frac{i}{2\pi\hbar\,t}\bigg)^{\frac{N+1}{2}},
\end{equation}
so we can assume
\begin{equation}\label{32}
  e^{\varphi(t)}=\bigg(\frac{i}{2\pi\hbar\,t}\bigg)^{\frac{N+1}{2}}e^{\lambda(t)},
\end{equation}
where the unknown function $\lambda(t)$ satisfies
\begin{equation}\label{33}
\lim_{t\rightarrow 0} \lambda(t)=0.
\end{equation}
The function $\mathcal{K}$ now has the form
\begin{eqnarray}\label{34}
  \mathcal{K}(\mathbf{y}'|\mathbf{y},t) &=& e^{\lambda(t)}\bigg(\frac{i}{2\pi\hbar t}\bigg)^{\frac{N}{2}}\,e^{-\frac{i}{2\hbar}[\mathbf{y}\cdot \mathbf{F}^{-1} \dot{\mathbf{F}}\cdot \mathbf{y}+\mathbf{y}'\cdot \mathbf{F}^{-1} \dot{\mathbf{F}}\cdot \mathbf{y}'-2\mathbf{y}'\cdot \mathbf{F}^{-1}\cdot \mathbf{y}]}.
\end{eqnarray}
To find $\lambda(t)$ we make use of the following identity
\begin{equation}\label{35}
  \delta(\mathbf{y}'-\mathbf{y})=\int d\mathbf{y}''\,\mathcal{K}(y'|y'',t)\,\mathcal{K}^{*}(y|y'',t),
\end{equation}
which can be easily checked using the definition of $\mathcal{K}$, Eq.(\ref{18-1}). By inserting Eq.(\ref{34}) and its complex conjugation into Eq.(\ref{35}) and doing the integral we will find
\begin{equation}\label{36}
  e^{\lambda(t)}=\frac{t^{\frac{N+1}{2}}}{\sqrt{|\det \mathbf{F}(t)|}}\,e^{i\theta},
\end{equation}
where $\theta$ is a real function that will be determined from a limiting case where the coupling constants are turned off ($g_1=\cdots=g_N=0$) and also the fact that the propagator should satisfy the Schr\"{o}dinger equation.

It should be noted that according to the definition Eq.(\ref{18-1}), the Feynman propagator has the following relation to the function $\mathcal{K}$
\begin{equation}\label{37}
  K(\mathbf{y},t;\mathbf{y}',0)=\la \mathbf{y},t|\mathbf{y}',0\ra=\la \mathbf{y}'|\mathbf{y},t\ra^{*}=\mathcal{K}^{*}(\mathbf{\mathbf{y}}'|\mathbf{\mathbf{y}},t),
\end{equation}
therefore, Feynman propagator is given by
\begin{eqnarray}\label{38}
 K(\mathbf{y},t;\mathbf{y}',0) &=& \frac{e^{-i\theta}}{\sqrt{|\det F(t)|}}\bigg(\frac{1}{2\pi i\hbar}\bigg)^{\frac{N+1}{2}}\,e^{\frac{i}{2\hbar}[\mathbf{y}\cdot \mathbf{F}^{-1} \dot{\mathbf{F}}\cdot \mathbf{y}+\mathbf{y}'\cdot \mathbf{F}^{-1} \dot{\mathbf{F}}\cdot \mathbf{y}'-2\mathbf{y}'\cdot \mathbf{F}^{-1}\cdot \mathbf{y}]}.\nonumber\\
\end{eqnarray}
Now we set $y'=0$ and require that $K(\mathbf{y},t;0,0)$ satisfy the  Schr\"{o}dinger equation
\begin{equation}\label{38-1}
  i\hbar\frac{\p K(\mathbf{y},t;0,0)}{\p t}=\bigg[\haf \sum_{\mu=0}^N \bigg(-\hbar^2\frac{\p^2}{\p y_\mu^2}+\omega_\mu^2 y_\mu^2\bigg)-
  \haf\sum_{\mu,\nu=0}^N y_\mu \Omega_{\mu\nu}^2 y_\nu\bigg]\,K(\mathbf{y},t;0,0),
\end{equation}
after spatial differentiations we set $y=0$ and by comparing both sides of Eq.(\ref{38-1}) we find that $\theta$ is a constant. To find the constant $\theta$, we turn off the coupling constants, ($g_1=g_2=\cdots=g_N=0$), and from consistency condition we should recover the quantum propagator of $N$ noninteracting oscillators. When the coupling constants are turned off, We have
\begin{eqnarray}\label{39}
  \mathbf{F}^{-1} (t) &=& \mbox{diag}\bigg(\frac{\omega_0}{\sin(\omega_0 t)},\frac{\omega_1}{\sin(\omega_1 t)},\cdots,\frac{\omega_N}{\sin(\omega_N t)}\bigg),\nonumber\\
  \dot{\mathbf{F}}(t) &=& \mbox{diag}(\cos(\omega_0 t),\cos(\omega_0 t),\cdots,\cos(\omega_0 t)),
\end{eqnarray}
Inserting Eqs.(\ref{39}) into Eq.(\ref{38}) we find
\begin{equation}\label{40}
  K(\mathbf{y},t;\mathbf{y}',0)=e^{-i\theta}\prod_{\mu=0}^N \sqrt{\frac{\omega_\mu}{2\pi i\hbar \sin(\omega_\mu t)}}
  \,e^{\frac{i \omega_\mu}{2\hbar\sin(\omega_\mu t)}\big[(y_\mu^2+{y'}^2_\mu)\cos(\omega_\mu t)-2y_\mu y'_\mu\big]},
\end{equation}
which is the propagator of $N$ noninteracting oscillators if we set $\theta=0$. Finally, we find the quantum propagator of oscillator-bath system as
\begin{eqnarray}\label{41}
  K(\mathbf{y},t;\mathbf{y}',0) &=& \frac{1}{\sqrt{\det F(t)}}\bigg(\frac{1}{2\pi i\hbar}\bigg)^{\frac{N+1}{2}}\,e^{\frac{i}{2\hbar}[\mathbf{y}\cdot \mathbf{F}^{-1} \dot{\mathbf{F}}\cdot \mathbf{y}+\mathbf{y}'\cdot \mathbf{F}^{-1} \dot{\mathbf{F}}\cdot \mathbf{y}'-2\mathbf{y}'\cdot \mathbf{F}^{-1}\cdot \mathbf{y}]}.\nonumber\\
\end{eqnarray}
%%%%%%%%%%%%%%%%%%%%%%%%%%%%%%%%%%%%%%%%%%%%%%%%%%%%%%%%%%%%%%%%%%%%%%%%%%%%%%%%%%%%%%%%%%%%%%%%%%%%%%%%%%%%%%%%%%%%%%%%%%%%%%%%%%%%%%%%%%
\section{Density matrix}
%%%%%%%%%%%%%%%%%%%%%%%%%%%%%%%%%%%%%%%%%%%%%%%%%%%%%%%%%%%%%%%%%%%%%%%%%%%%%%%%%%%%%%%%%%%%%%%%%%%%%%%%%%%%%%%%%%%%%%%%%%%%%%%%%%%%%%%%%%
\noindent In this section we will find the density matrix for the oscillator-bath system using the explicit form of the quantum propagator Eq.(\ref{41}) of the combined system. If we denote the evolution operator by $\hat{U}(t)$ then the density matrix at time $t$ can be obtained from the initial density matrix at $t=0$ as
\begin{equation}\label{42}
  \hat{\rho}(t)=\hat{U(}t) \hat{\rho}(0) \hat{U}^{\dag}(t),
\end{equation}
in position representation we have
\begin{eqnarray}\label{43}
  \rho(\mathbf{y},\mathbf{y}';t) &=& \la \mathbf{y}|\rho (t)|\mathbf{y}'\ra \nonumber\\
   &=& \int d\mathbf{y}_1 d\mathbf{y}_2\,\la \mathbf{y}|\hat{U}(t)|\mathbf{y}_1\ra\la \mathbf{y}_1|\rho (0)|\mathbf{y}_2\ra\la \mathbf{y}_2 |\hat{U}^{\dag}(t)|\mathbf{y}'\ra,\nonumber \\
   &=& \int d\mathbf{y}_1 d\mathbf{y}_2\, K(\mathbf{y},t;\mathbf{y}_1,0)\rho(\mathbf{y}_1,\mathbf{y}_2;0) K^{*}(\mathbf{y}',t;\mathbf{y}_2,0).
\end{eqnarray}
We can assume an arbitrary initial state for oscillator-bath system, but for simplicity we assume that the initial state is a product state as
\begin{equation}\label{44}
  \rho(\mathbf{y}_1,\mathbf{y}_2;0)=\rho_{red} (y_{10},y_{20};0)\otimes \rho_B (\vec{y}_1,\vec{y}_2;0),
\end{equation}
where $\mathbf{y}_1=(y_{10},\vec{y}_1)$ and $\mathbf{y}_2=(y_{20},\vec{y}_2)$. To find the reduced density matrix of the main oscillator, we should take trace over the degrees of freedom of the bath oscillators. Straightforward calculations lead to
\begin{eqnarray}\label{46}
  \rho_{red} (y_0,y'_0;t) &=& \frac{1}{|\det F|}\frac{1}{(2\pi\hbar)^{N+1}}\int dy_{01} dy_{02}\,e^{\frac{ia(y_0^2-{y'_0}^2)}{2\hbar}}\nonumber\\
  &\cdot & e^{\frac{ia(y^2_{01}-y^2_{02})}{2\hbar}-\frac{ib(y_0 y_{01}-y_0' y_{02})}{\hbar}}\rho_{red}(y_{01},y_{02};0)\nonumber\\
  &\cdot & \,I(y_0,y'_0;y_{01},y_{02}),
\end{eqnarray}
where
\begin{eqnarray}\label{I}
  I(y_0,y'_0;y_{01},y_{02}) &=& \int d\vec{y}\,e^{\frac{i}{\hbar}\sum\limits_{k=1}^N\big[(y_0-y'_0)B_k-(y_{01}-y_{02})C_k\big]y_k}\int d\vec{y}_1 d\vec{y}_2\,\rho_B (\vec{y}_1,\vec{y}_2;0)\nonumber\\
  &\cdot & \,e^{\frac{i}{2\hbar}\sum\limits_{k,l=1}^N y_{1k}A_{kl}y_{1l}}\,e^{\frac{i}{\hbar}\sum\limits_{k=1}^N \big[y_{01}B_k-y_0 C_k -\sum\limits_{l=1}^N D_{kl}y_l\big] y_{1k}}\nonumber\\
  &\cdot & e^{\frac{-i}{2\hbar}\sum\limits_{k,l=1}^N y_{2k}A_{kl}y_{2l}}e^{\frac{-i}{\hbar}\sum\limits_{k=1}^N
  \big[y_{02}B_k-y'_0 C_k -\sum\limits_{l=1}^N D_{kl}y_l\big]y_{2k}}.
\end{eqnarray}
The Eq. (\ref{46}) can be rewritten as
\begin{equation}\label{J1}
 \rho_{red} (y_0,y'_0;t) = \int dy_{01} dy_{02}\,J(y_0,y'_0;t|y_{01},y_{02})\,\rho_{red}(y_{01},y_{02};0),
\end{equation}
where
\begin{eqnarray}\label{J2}
J(y_0,y'_0;t|y_{01},y_{02}) &=& \frac{1}{|\det F|}\frac{1}{(2\pi\hbar)^{N+1}}\,e^{\frac{ia(y_0^2-{y'_0}^2)}{2\hbar}}\,e^{\frac{ia(y^2_{01}-y^2_{02})}{2\hbar}-\frac{ib(y_0 y_{01}-y_0' y_{02})}{\hbar}}\nonumber\\
& \times & I(y_0,y'_0;y_{01},y_{02}),
\end{eqnarray}
The function $J(y_0,y'_0;t|y_{01},y_{02})$, which can be interpreted as a reduced kernel, has been expressed in path integral language in terms of the Feynman-Vernon influence functional \cite{Path-0,Path-1,Path-4}. Here we have obtained the reduced kernel in terms of the quadratic integrals. The time dependent functions ($a, b$), vectors ($C_k, B_k$) and matrices ($A_{kl}, D_{kl}$) are defined by
\begin{eqnarray}\label{defs}
  a(t) &=& (F^{-1}\dot{F})_{00},\nonumber\\
  b(t) &=& (F^{-1})_{00},\nonumber\\
  C_k (t) &=& (F^{-1})_{k0}=(F^{-1})_{0k},\nonumber\\
  B_k (t) &=& (F^{-1}\dot{F})_{0k}=(F^{-1}\dot{F})_{k0},\nonumber\\
  A_{kl} &=& (\mathbf{A})_{kl}=(F^{-1}\dot{F})_{kl},\nonumber\\
  D_{kl} &=& (\mathbf{D})_{kl}=(F^{-1})_{kl}=(F^{-1})_{lk},
\end{eqnarray}
which can be rewritten more compactly in matrix form as
\begin{equation}\label{FF}
  F^{-1}\dot{F}=\left(
                  \begin{array}{cc}
                    a & \mathbf{B}^T \\
                    \mathbf{B} & \mathbf{A} \\
                  \end{array}
                \right),\,\,\,\, F^{-1}=\left(
                  \begin{array}{cc}
                    b & \mathbf{C}^T \\
                    \mathbf{C} & \mathbf{D} \\
                  \end{array}
                \right),
\end{equation}
\begin{equation}\label{CB}
  B^T=[B_1, B_2,\cdots,B_N],\,\,\,C^T=[C_1, C_2,\cdots,C_N].
\end{equation}
Let the initial state of the bath be a thermal state given by
\begin{eqnarray}\label{ex-1}
  \rho_B (\vec{y}_1,\vec{y}_2;0) &=& \bigg(\prod_{k=1}^N\sqrt{\frac{\omega_k}{2\pi\hbar\sinh(\beta\hbar\omega_k)}}\bigg)\nonumber\\
  &\times & e^{-\sum\limits_{k=1}^N \frac{\omega_k}{2\hbar\sinh(\beta\hbar\omega_k)}\big[(y_{1k}^2+y_{2k}^2)\cosh(\beta\hbar\omega_k)-2y_{1k}y_{2k}\big]},
\end{eqnarray}
then, the integrals over $\vec{y}_1$, $\vec{y}_2$ and $\vec{y}$ in Eq.(\ref{I}), will be Gaussian type integrals and can be obtained using the generic formula \cite{Zin}
\begin{equation}\label{zin}
  \int d\vec{x}\,e^{-\frac{1}{2}\sum\limits_{k,l=1}^N x_k \Gamma_{kl} x_l+\sum\limits_{k=1}^N j_k x_k}=(2\pi)^{N/2}(\det \Gamma)^{-1/2}
  e^{\frac{1}{2}\sum\limits_{k,l=1}^N j_k \Gamma^{-1}_{kl} j_l},
\end{equation}
where $\Gamma$ is a positive, symmetric matrix.
\subsection{Master equation}
\noindent The main ingredient quantity in open quantum system theory is the master equation. To find the master equation satisfied by the reduced density matrix $\rho_{red}$, we insert the initial bath state Eq.(\ref{ex-1}) into Eq.(\ref{I}) and take the integrals over $\vec{y}$, $\vec{y}_1$ and $\vec{y}_2$, after straightforward but tedious calculations we will find the following expression for the reduced kernel defined in Eq.(\ref{J2})
\begin{eqnarray}\label{J}
  J(y_0,y'_0;y_{01},y_{02}) &=& \frac{b_3}{2\pi}\,e^{ib_1 X\xi+ib_2 X_0\xi-ib_3 X\xi_0-ib_4 X_0 \xi_0}\nonumber\\
  &\times & e^{-a_{11} \xi^2-a_{12}\xi\xi_0-a_{22}\xi_0^2},
\end{eqnarray}
where for later convenience, we have chosen the same notation for the time-dependent coefficients $b_k (t)$ and $a_{ij} (t)$ introduced by Paz in \cite{Paz} following the path integral technique. These coefficients can be obtained in terms of the functions given by Eqs.(\ref{defs}) or in terms of the environment properties described in \cite{Paz}. Following the same process described by Paz in \cite{Paz}, we recover the master equation $(\hbar=1)$ for the reduced density matrix as
\begin{eqnarray}\label{Master}
 && i\frac{\partial \rho_{red} (y_0,y'_0,t)}{\partial t}=\la y_0|[H_{ren},\rho_{red}]|y'_0\ra-i\gamma(t)(y_0-y'_0)(\frac{\partial }{\partial y_0}-\frac{\partial}{\partial y'_0})\rho_{red}(y_0,y'_0,t),\nonumber\\
 && -i D(t) (y_0-y'_0)^2\,\rho_{red}(y_0,y'_0,t)+f(t) (y_0-y'_0) (\frac{\partial }{\partial y_0}+\frac{\partial}{\partial y'_0})\rho_{red}(y_0,y'_0,t),\nonumber
\end{eqnarray}
where $H_{ren}$ is the renormalized Hamiltonian of the main oscillator with the renormalized frequency $\omega_{ren} (t)$. To find the connection between the functions $\omega_{ren} (t), \gamma(t), D(t), f(t)$ and coefficients $b_k (t),\,a_{ij}(t)$, the interested reader is referred to \cite{Paz}.
%%%%%%%%%%%%%%%%%%%%%%%%%%%%%%%%%%%%%%%%%%%%%%%%%%%%%%%%%%%%%%%%%%%%%%%%%%%%%%%%%%%%%%%%%%%%%%%%%%%%%%%%%%%%%%%%%%%%%%%%%%%%%%%%%%%%%%%%%%
\section{Thermal Equilibrium: fixed point}
%%%%%%%%%%%%%%%%%%%%%%%%%%%%%%%%%%%%%%%%%%%%%%%%%%%%%%%%%%%%%%%%%%%%%%%%%%%%%%%%%%%%%%%%%%%%%%%%%%%%%%%%%%%%%%%%%%%%%%%%%%%%%%%%%%%%%%%%%%
\noindent In the equilibrium state, the density matrix of oscillator-bath system can be obtained from the quantum propagator using the correspondence between quantum propagator and partition function as
\begin{equation}\label{P1}
  \rho(y_0,\vec{y};y'_0,\vec{y}',\beta)=\frac{1}{Z(\beta)}\,K(y_0,\vec{y},-i\hbar\beta;y'_0,\vec{y}',0),
\end{equation}
where $\beta=1/\kappa_B T$ is the inverse of temperature and $\kappa_B$ is Boltzmann constant. The function $Z(\beta)$ is the total partition function
\begin{eqnarray}\label{P2}
  Z(\beta) &=& \int dy_0 d\vec{y}\,K(y_0,\vec{y},-i\hbar\beta;y_0,\vec{y},0),\nonumber\\
  &=& \frac{1}{2^{\frac{N+1}{2}}}\frac{1}{\sqrt{\det (\dot{F}-\mathbb{I})}}\bigg|_{t=-i\hbar\beta},
\end{eqnarray}
and $\mathbb{I}$ is a $N$-dimensional unit matrix.

The reduced density matrix of the oscillator is obtained by integrating out the bath degrees of freedom as
\begin{eqnarray}\label{p3}
  \rho_{red}(y_0,y'_0;\beta) &=& \int d\vec{y}\,K(y_0,\vec{y},-i\hbar\beta;y'_0,\vec{y},0),\nonumber\\
   &=& \sqrt{\frac{\det (\dot{F}-\mathbb{I})}{i\pi\hbar\det F\det(\mathbf{A}-\mathbf{D})}}\,e^{\frac{i}{2\hbar}\big[({y_0}^2+{y'_0}^2)(a-\frac{\eta}{2})-2 y_0 y'_0 (b+\frac{\eta}{2})\big]},\nonumber\\
\end{eqnarray}
where
\begin{equation}\label{p4}
  \eta=\sum_{k,l=1}^N(B_k-C_k)(\mathbf{A}-\mathbf{D})^{-1}_{kl} (B_l-C_l)|_{t=-i\hbar\beta}.
\end{equation}
From Eq.(\ref{FF}) we have
\begin{equation}\label{p5}
  F^{-1}(\dot{F}-\mathbb{I})=\left(
                               \begin{array}{cc}
                                 a-b & \mathbf{B}^T-\mathbf{C}^T \\
                                 \mathbf{B}-\mathbf{C} & \mathbf{A}-\mathbf{D} \\
                               \end{array}
                             \right),
\end{equation}
by making use of the identity \cite{Matrix}
\begin{eqnarray}\label{p6}
  \det[F^{-1}(F-I)] &=& \det(\mathbf{A}-\mathbf{D})\det[a-b-\underbrace{(\mathbf{B}^T-\mathbf{C}^T)(\mathbf{A}-\mathbf{D})^{-1}(\mathbf{B}-\mathbf{C})}_{\eta}],\nonumber\\
  &=& \det(\mathbf{A}-\mathbf{D})(a-b-\eta),
\end{eqnarray}
Eq.(\ref{p3}) can be rewritten as
\begin{equation}\label{p7}
  \rho_{red}(y_0,y'_0;\beta)=\sqrt{\frac{a-b-\eta}{i\hbar\pi}}\,e^{\frac{i}{2\hbar}\big[({y_0}^2+{y'_0}^2)(a-\frac{\eta}{2})-2 y_0 y'_0 (b+\frac{\eta}{2})\big]}.
\end{equation}
From Eq.(\ref{p7}) we find the thermal mean square of position and momentum as
\begin{eqnarray}
  \la y_0^2 \ra &=& \frac{i\hbar}{2(a-b-\eta)}\bigg|_{t=-i\hbar\beta},\nonumber \\
  \la p_0^2 \ra &=& -i\hbar\frac{a+b}{2}\bigg|_{t=-i\hbar\beta},
\end{eqnarray}
therefore,
\begin{equation}\label{p8}
  \rho_{red}(y_0,y'_0;\beta)=\frac{1}{\sqrt{2\pi\la y_0^2 \ra}}\,e^{-\frac{\la p_0^2 \ra}{2\hbar^2}(y_0-y_0')^2-\frac{1}{8\la y_0^2 \ra}(y_0+y_0')^2},
\end{equation}
for another derivation, see \cite{Weiss}.
%%%%%%%%%%%%%%%%%%%%%%%%%%%%%%%%%%%%%%%%%%%%%%%%%%%%%%%%%%%%%%%%%%%%%%%%%%%%%%%%%%%%%%%%%%%%%%%%%%%%%%%%%%%%%%%%%%%%%%%%%%%%%%%%%%%%%%%%%%%%%%%%%%%%%%%%%%%%%%%%
%%%%%%%%%%%%%%%%%%%%%%%%%%%%%%%%%%%%%%%%%%%%%%%%%%%%%%%%%%%%%%%%%%%%%%%%%%%%%%%%%%%%%%%%%%%%%%%%%%%%%%%%%%%%%%%%%%%%%%%%%%%%%%%%%%%%%%%%%%%%%%%%%%%%%%%%%%%%%%%%
\section{Main oscillator interacts with an external field}
%%%%%%%%%%%%%%%%%%%%%%%%%%%%%%%%%%%%%%%%%%%%%%%%%%%%%%%%%%%%%%%%%%%%%%%%%%%%%%%%%%%%%%%%%%%%%%%%%%%%%%%%%%%%%%%%%%%%%%%%%%%%%%%%%%%%%%%%%%
\noindent Now assume that the main oscillator is under the influence of an external classical field $f(t)$. In this case the total Lagrangian is written as
\begin{equation}\label{47}
  L=\haf \sum_{\mu=0}^N (\dot{Y}^2_\mu-\omega_\mu^2 Y_\mu^2)+\haf\sum_{\mu,\nu=0}^N Y_\mu \Omega_{\mu\nu}^2 Y_\nu-f(t) Y_0,
\end{equation}
and the corresponding Hamiltonian is
\begin{equation}\label{48}
  H=\haf \sum_{\mu=0}^N (P^2_\mu+\omega_\mu^2 Y_\mu^2)-\haf\sum_{\mu,\nu=0}^N Y_\mu \Omega_{\mu\nu}^2 Y_\nu+f(t) Y_0.
\end{equation}
Note that the Hamiltonian is now time-dependent and we can not use Eqs.(\ref{24},\ref{25}). In this case, we can find another partial differential equation satisfied by $\mathcal{K}(y'|y,t)$ as follows. From Heisenberg equations of motion we find
\begin{equation}\label{48-1}
   \ddot{\hat{Y}}_\mu +\omega_\mu^2 \hat{Y}_\mu-\sum_\nu \Omega_{\mu\nu}^2 \hat{Y}_\nu=-f(t)\,\delta_{\mu 0}.
\end{equation}
The Green tensor corresponding to Eq.(\ref{48-1}) is defined by
\begin{equation}\label{48-2}
  \sum_\nu\bigg(\big[\p^2_t +\omega_\mu^2\big]\delta_{\mu\nu}-\Omega^2_{\mu\nu}\bigg)G_{\nu\alpha} (t-t')=\delta_{\mu\alpha}\,\delta(t-t').
\end{equation}
By making use of Laplace transform and definitions Eqs.(\ref{9},\ref{12}), we find the retarded Green tensor as
\begin{equation}\label{48-3}
  G_{\mu\nu} (t-t')=F_{\mu\nu} (t-t'),
\end{equation}
and the position and momentum operators are respectively given by
\begin{eqnarray}\label{49}
&&  \hat{Y}_\mu (t)=\sum_{\nu} \big[\dot{F}_{\mu\nu} (t) \hat{Y}_\nu (0)+F_{\mu\nu} (t) \hat{P}_\mu (0)\big]-R_\mu (t),\nonumber\\
&&  \hat{P}_\mu =\dot{\hat{Y}}_\mu=\sum_{\nu} \big[\ddot{F}_{\mu\nu} (t) \hat{Y}_\nu (0)+\dot{F}_{\mu\nu} (t) \hat{P}_\mu (0)\big]-\dot{R}_\mu (t),
\end{eqnarray}
where we defined
\begin{equation}\label{50}
  R_\mu (t)=\int_0^t dt'\,F_{\mu 0} (t-t') f(t').
\end{equation}
We can rewrite the identity
\begin{equation}\label{51}
  \hat{P}_\mu (t)=\hat{U}^{\dag}(t) \hat{P}_\mu (0) \hat{U}(t),
\end{equation}
as
\begin{equation}\label{52}
  \hat{P}_\mu (t) \hat{U}^{\dag}(t)=\hat{U}^{\dag}(t) \hat{P}_\mu (0),
\end{equation}
then
\begin{equation}\label{53}
  \la y'|\hat{P}_\mu (t) \hat{U}^{\dag}(t)|y\ra=\la y'|\hat{U}^{\dag}(t) \hat{P}_\mu (0)|y\ra.
\end{equation}
By inserting the momentum operator from the second line of Eqs.(\ref{49}) into Eq.(\ref{53}), we easily find
\begin{equation}\label{54}
  \sum_{\nu}\bigg(\ddot{F}_{\mu\nu} (t)\,y'_\nu-i\hbar\,\dot{F}_{\mu\nu} (t)\,\frac{\p}{\p y'_\nu}-\dot{g}_\mu \bigg)
  \,\mathcal{K}(\mathbf{y}'|\mathbf{y},t)=y_\mu\,\mathcal{K}(\mathbf{y}'|\mathbf{y},t).
\end{equation}
By making use of Eqs.(\ref{18},\ref{29},\ref{35},\ref{54}), and following the same process as we did in Sec.III, we will find
\begin{eqnarray}\label{55}
  K^{(f)}(\mathbf{y},t;\mathbf{y}',0) &=& \frac{e^{-i\zeta(t)}}{\sqrt{|\det F(t)|}}\bigg(\frac{1}{2\pi i\hbar}\bigg)^{\frac{N+1}{2}}\,e^{\frac{i}{2\hbar}\big[\mathbf{y}\cdot \mathbf{F}^{-1} \dot{\mathbf{F}}\cdot \mathbf{y}+\mathbf{y}'\cdot \mathbf{F}^{-1} \dot{\mathbf{F}}\cdot \mathbf{y}'-2\mathbf{y}'\cdot \mathbf{F}^{-1}\cdot \mathbf{y}\big]}\nonumber\\
  &\times & e^{-\frac{i}{\hbar}\,\big[\mathbf{y}'\cdot \mathbf{F}^{-1}\cdot \mathbf{R} +\mathbf{y}\cdot \mathbf{F}^{-1}\cdot\check{\mathbf{R}}\big]},
\end{eqnarray}
where we have defined $\check{\mathbf{R}}$ as
\begin{equation}\label{56}
   \check{R}_\mu (t)=\int_0^{t} dt'\,F_{\mu 0}(t') f(t'),
\end{equation}
and the function $\zeta(t)$ can be determined from the Schr\"{o}dinger equation
\begin{eqnarray}\label{57}
 &&  i\hbar\frac{\p K^{(f)}(\mathbf{y},t;0,0)}{\p t}\bigg|_{y=0}=\nonumber\\
 && \bigg[\haf \sum_{\mu=0}^N \bigg(-\hbar^2\frac{\p^2}{\p y_\mu^2}+\omega_\mu^2 y_\mu^2\bigg)-
  \haf\sum_{\mu,\nu=0}^N y_\mu \Omega_{\mu\nu}^2 y_\nu +f(t) y_0\bigg]\,K^{(f)}(\mathbf{y},t;0,0)\bigg|_{y=0},\nonumber\\
\end{eqnarray}
as
\begin{eqnarray}\label{58}
  \zeta(t) &=& \frac{1}{2\hbar}\int_0^t ds\,\check{\mathbf{R}}(s)\cdot \mathbf{F}^{-2} (s)\cdot \check{\mathbf{R}}(s),\nonumber\\
           &=& \frac{1}{\hbar}\int_0^t ds\,\int_0^{s} du\,f(s)\bigg[\frac{\sin(\sqrt{B}u)\sin[\sqrt{B}(t-s)]}{\sqrt{B}\sin(\sqrt{B}t)}\bigg]_{00}f(u).
\end{eqnarray}
Finally, the quantum propagator for oscillator-bath system under the influence of an external classical force on the main oscillator, is obtained as
 \begin{eqnarray}\label{59}
  K^{(f)}(\mathbf{y},t;\mathbf{y}',0) &=& \frac{1}{\sqrt{|\det F(t)|}}\bigg(\frac{1}{2\pi i\hbar}\bigg)^{\frac{N+1}{2}}\,e^{\frac{i}{2\hbar}\big[\mathbf{y}\cdot \mathbf{F}^{-1} \dot{\mathbf{F}}\cdot \mathbf{y}+\mathbf{y}'\cdot \mathbf{F}^{-1} \dot{\mathbf{F}}\cdot \mathbf{y}'-2\mathbf{y}'\cdot \mathbf{F}^{-1}\cdot \mathbf{y}\big]}\nonumber\\
  &\times & e^{-\frac{i}{\hbar}\,\big[\mathbf{y}'\cdot \mathbf{F}^{-1}\cdot \mathbf{R} +\mathbf{y}\cdot \mathbf{F}^{-1}\cdot\check{\mathbf{R}}\big]}\,e^{-\frac{i}{2\hbar}\int_0^t ds\,\check{\mathbf{R}}(s)\cdot \mathbf{F}^{-2} (s)\cdot \check{\mathbf{R}}(s)}.
\end{eqnarray}
%%%%%%%%%%%%%%%%%%%%%%%%%%%%%%%%%%%%%%%%%%%%%%%%%%%%%%%%%%%%%%%%%%%%%%%%%%%%%%%%%%%%%%%%%%%%%%%%%%%%%%%%%%%%%%%%%%%%%%%%%%%%%%%%%%%%%%%%%%%%%%%%%%%%%%%%%%%%%%
\subsection{A generalization: generating function}
%%%%%%%%%%%%%%%%%%%%%%%%%%%%%%%%%%%%%%%%%%%%%%%%%%%%%%%%%%%%%%%%%%%%%%%%%%%%%%%%%%%%%%%%%%%%%%%%%%%%%%%%%%%%%%%%%%%%%%%%%%%%%%%%%%%%%%%%%%%%%%%%%%%%%%%%%%%%%%
\noindent We can generalize the Lagrangian Eq.(\ref{47}) as
\begin{equation}\label{60}
  L=\haf \sum_{\mu=0}^N (\dot{Y}^2_\mu-\omega_\mu^2 Y_\mu^2)+\haf\sum_{\mu,\nu=0}^N Y_\mu \Omega_{\mu\nu}^2 Y_\nu-\sum_{\mu=0}^N f_{\mu}(t) Y_\mu,
\end{equation}
in this case the quantum propagator is given by Eq.(\ref{50}) but now the definitions Eqs.(\ref{50},\ref{56}) have to be replaced by the new definitions
\begin{eqnarray}\label{61}
  R_\mu (t) &=& \int_0^t dt'\,F_{\mu\nu} (t-t') f_{\nu} (t'),\nonumber\\
  \check{R}_\mu (t) &=& \int_0^t dt'\,F_{\mu\nu} (t') f_{\nu} (t').
\end{eqnarray}
The path integral representation of quantum propagator Eq.(\ref{59}) is \cite{Greiner}
\begin{equation}\label{62}
  K^{(f)}(\mathbf{y},t;\mathbf{y}',0)=\int d[\mathbf{x}]\,e^{\frac{i}{\hbar}\int_0^t d\tau\,L},
\end{equation}
where $L$ is the Lagrangian Eq.(\ref{60}). Having the closed form expression Eq.(\ref{59}), we can find ordered correlation functions among position operators of the oscillator-bath system. In this case, the external source $f_\mu (t)$ is an auxiliary force that should be set zero at the end of functional derivatives \cite{Greiner}, we have
\begin{eqnarray}\label{63}
&&  \la \mathbf{y},t|\hat{T}[\hat{Y}_{\mu_1}(t_1) \hat{Y}_{\mu_2}(t_2)\cdots \hat{Y}_{\mu_N}(t_N)|\mathbf{y}',0 \ra = \nonumber\\
\vspace{1.cm}
&&  \frac{\int D[\mathbf{x}]\,y_{\mu_1}(t_1) y_{\mu_2}(t_2)\cdots y_{\mu_N}(t_N)\,e^{\frac{i}{\hbar}\int_0^t d\tau\,L}}{\int D[\mathbf{x}]\,e^{\frac{i}{\hbar}\int_0^t d\tau\,L}}=
  \nonumber \\
&& \frac{(i\hbar)^N}{K^{(0)}(\mathbf{y},t;\mathbf{y}',0)} \frac{\delta^{N}}{\delta f_{\mu_1} (t_1)\cdots \delta f_{\mu_N} (t_N)} K^{(f)}(\mathbf{y},t;\mathbf{y}',0)\bigg|_{f=0},
\end{eqnarray}
where $\hat{T}$ is a time ordering operator acting on bosonic operators as
\begin{equation}\label{64}
  \hat{T}(\hat{A}(t)\hat{B}(t')=\left\{
                            \begin{array}{ll}
                              \hat{A}(t)\hat{B}(t'), & \hbox{$t>t'$;} \\
                              \hat{B}(t')\hat{A}(t), & \hbox{$t'>t$.}
                            \end{array}
                          \right.
\end{equation}
%%%%%%%%%%%%%%%%%%%%%%%%%%%%%%%%%%%%%%%%%%%%%%%%%%%%%%%%%%%%%%%%%%%%%%%%%%%%%%%%%%%%%%%%%%%%%%%%%%%%%%%%%%%%%%%%%%%%%%%%%%%%%%%%%%%%%%%%%%%%%%%%%%%%%%%%%%%%%%
\section{Conclusions}
%%%%%%%%%%%%%%%%%%%%%%%%%%%%%%%%%%%%%%%%%%%%%%%%%%%%%%%%%%%%%%%%%%%%%%%%%%%%%%%%%%%%%%%%%%%%%%%%%%%%%%%%%%%%%%%%%%%%%%%%%%%%%%%%%%%%%%%%%%%%%%%%%%%%%%%%%%%%%%
\noindent Using elementary quantum mechanical calculations and basic properties of quantum propagators, an alternative derivation of exact quantum propagator for the oscillator-bath system was introduced.
The method compared to other methods to derive quantum propagator of an oscillator-bath system with linear interaction or generally quadratic Lagrangians, was easier to apply and in particular, compared to path integral approach, there was no need to introduce more advanced mathematical notions like infinite integrations, operator determinant and Weyl ordering. From quantum propagator, a closed form density matrix describing the combined oscillator-bath system was obtained from which reduced density matrix could be derived. The problem was generalised to the case where the main oscillator was under the influence of an external classical source. By introducing auxiliary classical fields the modified quantum propagator or generating functional of position correlation functions was found.

The basic ingredient of the approach was a symmetric time-independent matrix $B$, which was dependent on natural frequencies of the bath oscillators and coupling constants. Therefore, from numerical or simulation point of view, the only challenge was finding the inverse of the matrix $B$ or equivalently diagonalizing it.

The efficiency of the method in determining the exact form of the quantum propagator for quadratic Lagrangians, inspired the idea of developing a perturbative approach to include non-quadratic Lagrangians too.
%%%%%%%%%%%%%%%%%%%%%%%%%%%%%%%%%%%%%%%%%%%%%%%%%%%%%%%%%%%%%%%%%%%%%%%%%%%%%%%%%%%%%%%%%%%%%%%%%%%%%%%%%%%%%%%%%%%%%%%%%%%%%%%%%%%%%%%%%%%%%%%%%%%%%%%%%%%%%%


\begin{thebibliography}{00}
%%%%%%%%%%%%%%%%%%%%%%%%%%%%%%%%%%%%%%%%%%%%%%%%%%%%%%%%%%%%%%%%%%%%%%%%%%%%%%%%%%%%%%%%%%%%%%%%%%%%%%%%%%%%%%%%%%%%%%%%%%%%%%%%%%%%%%%%%%%%%%%%%%%%%%%%%%%%%%
\bibitem{Propagator-1} W. H. Dickhoff and D. V. Neck, Many-Body Theory Exposed! Propagator description of
quantum mechanics in many-body systems, World Scientific, 2005.

\bibitem{Propagator-2} J. Linderberg and Y. \"{O}hrn, Propagators in quantum chemistry, John Wiley,  2004.

\bibitem{Feynman-1} R. P. Feynman, A. R. Hibbs, D. F. Styer, Quantum Mechanics and Path Integrals, Emended Ed., McGrawHill, 2005.

\bibitem{Kleinert} H. Kleinert, Path Integrals in Quantum Mechanics, Statistics,
Polymer Physics, and Financial Markets, 5th ed., World
Scientific, 2009.

\bibitem{Zin} J. Zinn-Justin, Path Integrals in Quantum Mechanics, Oxford, 2005.

\bibitem{Path-0} R. Feynman, F. Vernon Jr., The theory of a general quantum system inter-
acting with a linear dissipative system, Annals of Physics 24 (1963) 118–173.

\bibitem{Path-1} A. O. Caldeira and A. J. Leggett, Path integral approach to quantum Brownian motion, Physica A 121 (1963) 587-616.

\bibitem{Path-2} V. Hakim and V. Ambegoakar, Quantum theory of a free particle interacting with a linearly dissipative environment, Phys. Rev. A 32 (1985) 423-434.

\bibitem{Path-3} C. Morais Smith and A. O. Caldeira, Generalized Feynman-Vernon approach to dissipative quantum systems, Phys. Rev. A 36 (1987) 3509-3511.

\bibitem{Path-4} H. Grabert, P. Schramm, and G.-L. Ingold, Quantum Brownian Motion: The functional integral approach, Phys. Rep. 168 (1988) 115-207.

\bibitem{Path-5} B. L. Hu, J. P. Paz, and Y. Zhang, Quantum Brownian motion in a general environment: Exact master equation
with nonlocal dissipation and colored noise, Phys. Rev. D 45 2843 (1992) 2843-2861.

\bibitem{Path-6} B. L. Hu and A Matacz, Quantum Brownian motion in a bath of parametric oscillators: A model for
system-field interactions, Phys. Rev. D 49 (1994) 6612-6635.

\bibitem{Greiner} W. Greiner and J. Reinhardt, Field Quantization, Springer-
Verlag, 1996.

\bibitem{Weiss} U. Weiss, Dissipative Quantum Systems, World Scientific, 1993.

\bibitem {Haake} F. Haake and R. Reibold, Strong damping and low-temperature anomalies for the harmonic oscillator, Phys. Rev. A 32 (1985) 2462-2475.

\bibitem {Ullersma} P. Ullersma, An exactly solvable model for Brownian motion: I. Derivation of the Langevin equation, Physica (Utrecht) 32 (1966) 27-55.

\bibitem{Paz} J. P. Paz, Decoherence in quantum Brownian motion, in Physical origins of time asymmetry, Edited by J. J. Halliwell, J. P\'{e}rez-Mercader and W. H. Zurek, Cambridge University Press, 1994.

\bibitem{Matrix} F. Zhang, Matrix Theory: Basic Results and Techniques, Springe-Verlag, 2010.

%%%%%%%%%%%%%%%%%%%%%%%%%%%%%%%%%%%%%%%%%%%%%%%%%%%%%%%%%%%%%%%%%%%%%%%%%%%%%%%%%%%%%%%%%%%%%%%%%%%%%%%%%%%%%%%%%%%%%%%%%%%%%%%%%%%%%%%%%%%%%%%%%%%%%%%%%%%%%%
\end{thebibliography}
\end{document}